# MODELES MULTIPARTICULAIRES DES MATERIAUX MULTICOUCHES M4_5n et M4_(2n+1)M POUR L'ANALYSE DES EFFETS DE BORD

## MULTIPARTICLE MODELS OF MULTILAYERED MATERIALS M4_5n AND M4_(2n+1)M FOR EDGE EFFECT ANALYSIS

Armelle CHABOT et Alain EHRLACHER[*]

[*]Ecole Nationale des Ponts et chaussées
Centre d'Enseignement et de Recherche en Analyse des Matériaux
6 et 8 avenue Blaise Pascal, Cité Descartes - Champs sur Marne
77455 Marne La Vallée Cedex 2 - France

## RESUME

Les Modélisations Multiparticulaires des Matériaux Multicouches (modèles M4) sont conçus pour analyser simplement les champs de contraintes 3D provoquant du délaminage ou de la fissuration transverse dans les stratifiés. Elles sont construites à partir de champs de contraintes 3D approchés, polynomiaux en z par couche, et de la formulation d'Hellinger-Reissner. Nous présentons ici deux de ces modèles: le M4_5n qui a une cinématique de plaques de Reissner par couche, soit 5n champs (n: nombre de couches), et le modèle M4_(2n+1)M qui a principalement une cinématique de membrane par couche, soit 2n+1 champs. Ce dernier modèle conduit à des solutions analytiques simples et à un concept d'effort linéique concentré au bord pouvant servir de base à un critère de délaminage. Nous illustrons donc quelques une de ses solutions.

## ABSTRACT

The Multiparticle Models of Multilayered Materials (models M4) are developed to analyze easily the 3D stress fields which produce delamination or transverse crackings in laminates. The M4 models are built from 3D approximate stress fields in z per layer and the Hellinger-Reissner's formulation. Here-in we express two of these models: the M4_5n model of which the kinematics, with 5n fields (n: number of layers), is the one of n Reissner's plates and, the M4_(2n+1)M model of membrane with 2n+1 fields. The latter gives simple analytic solutions and a concentrated edge force per unit length that could be a natural support for the criteria of delamination. We illustrate this concept by the analytic resolution of $(0°, 90°)_s$ in traction and give some results for more complex stackings.

MODELES - STRATIFIES - MULTIPARTICULAIRE - EFFETS DE BORD - DELAMINAGE
MODELS - LAMINATES - MULTIPARTICLE - EDGE EFFECTS - DELAMINATION

## INTRODUCTION

Les phénomènes d'endommagement qui précèdent la ruine des structures en matériaux composites sont principalement pilotés par les efforts à l'interface entre les couches. Les théories classiques de Love-Kirchhoff ou de Reissner-Mindlin permettent une bonne description des déformations globales et des champs de contrainte de la plaque, sauf au voisinage des bords, jusqu'à une distance de 3 ou 4 fois l'épaisseur, là où la concentration de contrainte est 3D, c'est à dire là où il y a effets de bord.

Depuis les premières modélisations numériques de Pipes et Pagano (1970) aux modélisations éléments finis très fines de Wang et Crossman (1977) ou de Raju et Crews (1981), plusieurs auteurs ont cherché à développer des modélisations simplifiées capable d'analyser ces concentrations de contraintes 3D. Parmi elles, l'utilisation d'une approximation du champ de contrainte à partir de champs en (x, y) définis par couche nous semble conduire à des modèles assez simples. Ce choix revient à retenir une cinématique généralisée définie elle aussi à partir de champs en (x, y) par couche. On est ainsi amené à considérer qu'en un point de la surface définissant géométriquement l'objet se trouvent n particules, n étant le nombre de couches. Ce sont les "modèles multiparticulaires".

Les Modélisations Multiparticulaires des Matériaux Multicouches (modèles M4) (Chabot, 1997) sont construites à partir de champs de contraintes tridimensionnels approchés écrits sous forme de polynômes de Legendre en z par couche et de la formulation d'Hellinger-Reissner (Reissner, 1950). Pagano (1978) a procédé ainsi pour construire son modèle local à 7n champs cinématiques. Deux modèles M4 semblent particulièrement intéressants car facilement utilisables pour un ingénieur de bureau d'études. Le premier de ces deux modèles est le M4_5n dont la cinématique, à 5n champs, est celle de n plaques de Reissner (une par couche). Le deuxième modèle, le M4_(2n+1)M (M :membrane), à 2n+1 champs, présente un avantage décisif puisqu'il conduit dans beaucoup de problèmes à des solutions analytiques simples (Naciri et al, 1998) et fait apparaître un effort concentré au bord, de type Dirac.

Nous présentons ici les équations de ces deux modèles multiparticulaires et illustrons sur des exemples de stratifiés $(0°, 90°)_s$, $(\pm\theta)_s$ et $(\theta, \theta-90°)_s$ en traction.

Nous supposons, par la suite qu'il n'y a pas de forces de volume.

**Notations:** $\alpha, \beta, \gamma$ et $\delta \in \{1,2\}$ ; $\underline{A}$, $\overline{\overline{A}}$ et $\overline{\overline{\overline{A}}}$ (respectivement $\tilde{B}$, $\tilde{\tilde{B}}$ et $\tilde{\tilde{\tilde{B}}}$) sont des tenseurs d'ordre 1, 2 et 4 tridimensionnels (respectivement dans le plan médian $(x,y)$ du multicouche). La ième couche occupe le domaine $\omega \times [h_i^-, h_i^+]$ $(h_i^+ = h_{i+1}^-)$ $(i \in [1, n])$, son épaisseur est $e_i = h_i^+ - h_i^-$, sa côte moyenne est $\overline{h}_i = \dfrac{h_i^+ + h_i^-}{2}$.

## EQUATIONS DU MODELE M4_5n

Pour le modèle à 5n champs, nous choisissons d'approximer les contraintes tridimensionnelles par des polynômes en z à l'aide des efforts de plaque de Reissner-Mindlin $\tilde{N}^i$ $\tilde{M}^i$ et $\tilde{Q}^i$ de la couche i, des efforts de cisaillement $\tilde{\tau}^{i,i+1}$ et d'arrachement $\nu^{i,i+1}$ de chaque interface entre les couches i-1,i et i,i+1. Ce champ de contrainte tridimensionnel approché est introduit dans la fonctionnelle d'Hellinger-Reissner. On en déduit les 5n déplacements généralisés du modèle. Ce sont les déplacements généralisés de plaque de Reissner-Mindlin $\tilde{U}^i, U_3^i$ et $\tilde{\Phi}^i$ par couche. Les stationnarités de la fonctionnelle donnent les équations du modèle.

Les équations d'équilibre du modèle M4_5n s'écrivent alors:

(1) $\quad N_{\alpha\beta,\beta}^i(x,y) + \left(\tau_\alpha^{i,i+1}(x,y) - \tau_\alpha^{i-1,i}(x,y)\right) = 0$ sur $\omega$

(2) $\quad Q_{\alpha,\alpha}^i(x,y) + \left(\nu^{i,i+1}(x,y) - \nu^{i-1,i}(x,y)\right) = 0$ sur $\omega$

(3) $\quad Q_\alpha^i(x,y) = M_{\alpha\beta,\beta}^i(x,y) + \dfrac{e_i}{2}\left(\tau_\alpha^{i-1,i}(x,y) + \tau_\alpha^{i,i+1}(x,y)\right)$ sur $\omega$

avec pour conditions aux limites ($\underline{T}^d(x,y,z)$ imposé sur $\delta\omega \times \left[h_1^-, h_n^+\right]$):

(4) $\quad \begin{aligned} \tilde{\tilde{N}}^i . \underline{n} &= \tilde{T}_d^i \\ \tilde{\tilde{M}}^i . \underline{n} &= \tilde{M}_d^i \\ \tilde{Q}^i . \underline{n} &= Q_{3d}^i \end{aligned}$ où: $\begin{cases} T_{d\alpha}^i = \displaystyle\int_{h_i^-}^{h_i^+} T_\alpha^d(x,y,z)dz \;;\; Q_{3d}^i = \displaystyle\int_{h_i^-}^{h_i^+} T_3^d(x,y,z)dz \\ M_{d\alpha}^i = \displaystyle\int_{h_i^-}^{h_i^+} (z-\bar{h}^i)T_\alpha^d(x,y,z)dz \end{cases}$ $(x,y) \in \delta\omega$

Dans l'énergie élastique écrite en contrainte, pour simplifier l'écriture du comportement, nous négligeons les termes en $\tau_{\alpha,\alpha}^{i,i+1}$ et les énergies couplant les efforts membranaires et les contraintes perpendiculaires aux couches. Cette dernière hypothèse est habituelle dans la plupart des théories de plaque. Notons $S_{jklm}^i$ $i \in [1,n]$ $j,k,l,m \in [1,3]^4$ les composantes du tenseur d'ordre 4 des souplesses du matériau de la couche i. Si $\underline{e}_3$ est une direction principale d'orthotropie, pour chaque couche les comportements de couche associés aux efforts membranaires, aux moments de flexion et aux efforts de cisaillement hors plan s'écrivent:

(5) $\quad \varepsilon_{\alpha\beta}^i(x,y) = \dfrac{1}{2}\left(U_{\alpha,\beta}^i(x,y) + U_{\beta,\alpha}^i(x,y)\right) = \dfrac{S_{\alpha\beta\gamma\delta}^i}{e^i} : N_{\gamma\delta}^i(x,y)$

(6) $\quad \chi_{\alpha\beta}^i(x,y) = \dfrac{1}{2}\left(\Phi_{\alpha,\beta}^i(x,y) + \Phi_{\beta,\alpha}^i(x,y)\right) = \dfrac{12}{e^{i^3}} S_{\alpha\beta\gamma\delta}^i : M_{\gamma\delta}^i(x,y)$

(7) $\quad \begin{aligned} d_{Q_\alpha}^i(x,y) &= \Phi_\alpha^i(x,y) + U_{3,\alpha}^i(x,y) \\ &= \left(\dfrac{6}{5e^i}\right) S_{Q_{\alpha\beta}}^i . Q_\beta^i(x,y) - \left(\dfrac{1}{10}\right) S_{Q_{\alpha\beta}}^i . \left(\tau_\beta^{i-1,i}(x,y) + \tau_\beta^{i,i+1}(x,y)\right) \end{aligned}$

où $\tilde{\tilde{\varepsilon}}^i, \tilde{\tilde{\chi}}^i$ et $\tilde{d}_Q^i$ sont respectivement les tenseurs plans des déformations membranaires, des rotations membranaires et des déformations liées aux efforts tranchants.

Les comportements associés aux efforts de cisaillement et d'arrachement d'interface s'écrivent:

$$D_\alpha^{i,i+1}(x,y) = \left(U_\alpha^{i+1}(x,y) - U_\alpha^i(x,y) - \frac{e^i}{2}\Phi_\alpha^i(x,y) - \frac{e^{i+1}}{2}\Phi_\alpha^{i+1}(x,y)\right)$$

$$(8) \quad = -\frac{1}{10}4S_{\alpha3\beta3}^i.Q_\beta^i(x,y) - \frac{1}{10}4S_{\alpha3\beta3}^{i+1}.Q_\beta^{i+1}(x,y) + \left(\frac{-e^i}{30}\right)4S_{\alpha3\beta3}^i.\tau_\beta^{i-1,i}(x,y)$$

$$+ \frac{2}{15}\left(e^i 4S_{\alpha3\beta3}^i + e^{i+1} 4S_{\alpha3\beta3}^{i+1}\right)\tau_\beta^{i,i+1}(x,y) + \left(\frac{-e^{i+1}}{30}\right)4S_{\alpha3\beta3}^{i+1}.\tau_\beta^{i+1,i+2}(x,y)$$

$$(9) \quad D_3^{i,i+1}(x,y) = U_3^{i+1}(x,y) - U_3^i(x,y) = \frac{9}{70}e^i S_{3333}^i \nu^{i-1,i}(x,y)$$

$$+ \frac{13}{35}\left(e^i S_{3333}^i + e^{i+1} S_{3333}^{i+1}\right)\nu^{i,i+1}(x,y) + \frac{9}{70}e^{i+1} S_{3333}^{i+1}\nu^{i+1,i+2}(x,y)$$

où $\tilde{D}^{i,i+1}$ et $D_3^{i,i+1}$ sont les déformations associées respectivement aux cisaillements d'interface et aux arrachements d'interface.

Notons que le comportement associé aux efforts tranchants et cisaillement d'interface couple 3 interfaces successives. Ceci implique que le couplage des efforts tranchants et des cisaillements d'interface via les équations (8) et (9) implique que, du point de vue du comportement, le modèle M4_5n ne peut être vu comme une simple superposition de n plaques de Reissner-Mindlin

## EQUATIONS DU MODELE M4_(2n+1)M

Le modèle à (2n+1) champs est construit à partir d'un champ de contrainte 3D approché, plus simple, qui n'utilise que les efforts membranaires $\tilde{\tilde{N}}^i$ de chaque couche et les cisaillements $\tilde{\tau}^{i,i+1}$ de chaque interface i,i+1. La même méthode que précédemment donne (2n+1) déplacements généralisés qui sont les déplacements dans le plan (x,y) $\tilde{U}^i(x,y)$ de chaque couche et un déplacement vertical commun à toutes les couches $W_3(x,y)$.

Les équations d'équilibre du modèle M4_(2n+1)M s'écrivent:

$$(10) \quad N_{\alpha\beta,\beta}^i(x,y) + \left(\tau_\alpha^{i,i+1}(x,y) - \tau_\alpha^{i-1,i}(x,y)\right) = 0 \text{ sur } \omega$$

$$(11) \quad \sum_{j=1}^n \left[\frac{e_j}{2}\left(\tau_{\alpha,\alpha}^{j-1,j}(x,y) + \tau_{\alpha,\alpha}^{j,j+1}(x,y)\right)\right] + \left(T_3^-(x,y) + T_3^+(x,y)\right) = 0 \text{ sur } \omega$$

où $T_3^-$ (respectivement $T_3^+$) est la composante normale du vecteur contrainte imposé sur la face externe inférieure (respectivement supérieure) du multicouche avec pour conditions aux limites:

$$(12) \quad \begin{aligned} \tilde{\tilde{N}}^i \cdot \underline{n} &= \tilde{T}_d^i \\ \left( \sum_{i=1}^{n} \frac{e^i}{2} \left( \tilde{\tau}^{i,i+1} + \tilde{\tau}^{i-1,i} \right) \right) \cdot \underline{n} &= Q_3^d \end{aligned} \quad \text{si on a} \begin{cases} T_{d\alpha}^i = \int_{h_i^-}^{h_i^+} T_\alpha^d(x,y,z) dz \\ Q_{3d}^i = \int_{h_i^-}^{h_i^+} T_3^d(x,y,z) dz \end{cases}$$

Les lois de comportements avec les mêmes simplifications que pour le modèle M4_5n s'écrivent:

$$(13) \quad \varepsilon_{\alpha\beta}^i(x,y) = \frac{1}{2}\left(U_{\alpha,\beta}^i(x,y) + U_{\beta,\alpha}^i(x,y)\right) = \frac{S_{\alpha\beta\gamma\delta}^i}{e^i} : N_{\gamma\delta}^i(x,y)$$

$$D_\alpha^{i,i+1}(x,y) = \left( U_\alpha^{i,i+1}(x,y) - U_\alpha^i(x,y) + \frac{e^i + e^{i+1}}{2} W_{3,\alpha}(x,y) \right)$$

$$(14) \quad = \left(\frac{e^i}{6}\right) 4S_{\alpha 3\beta 3}^i \cdot \tau_\beta^{i-1,i}(x,y) + \frac{1}{3}\left(e^i 4S_{\alpha 3\beta 3}^i + e^{i+1} 4S_{\alpha 3\beta 3}^{i+1}\right) \tau_\beta^{i,i+1}(x,y)$$

$$+ \left(\frac{e^{i+1}}{6}\right) 4S_{\alpha 3\beta 3}^{i+1} \cdot \tau_\beta^{i+1,i+2}(x,y)$$

Ce modèle apparaît très similaire au modèle Shear Lag Analysis popularisé par Garett et Bailey (1977) excepté le fait que la cinématique soit définie à l'aide de 2n+1 champs (le modèle prend en compte la flèche globale $W_3$ du multicouche) et que le "comportement d'interface" couple trois interfaces consécutives.

## PLAQUE (0°,90°)$_s$ EN TRACTION

Nous nous intéressons dans cette partie à l'étude des effets de bord d'un empilement (0°,90°)$_s$ chargé en traction suivant la direction des fibres à 0°. Pour pouvoir comparer les prévisions de nos modèles avec celles données dans la littérature, nous reprenons les mêmes données que celles retenues par Wang et Crossman (1977) et Pagano (1978). L'empilement des quatre couches se fait suivant z. Nous numérotons les couches de 1 à 4 du bas vers le haut. Les couches 1 et 4 sont des couches à 0° (par rapport à Ox) et les couches 2 et 3 à 90°. Le domaine occupé par le multicouche est $x \in [-l,l]$ $y \in [-b,b]$ et $z \in [-2e, 2e]$ où e est l'épaisseur de chaque couche. Dans le problème étudié par Wang et Crossman et Pagano la largeur b=8e, la longueur l est très grande.

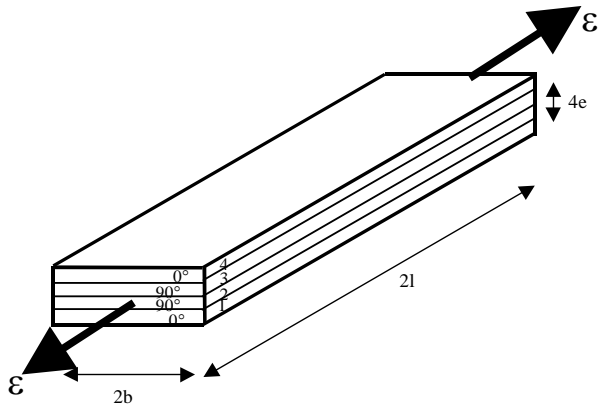

$E_{11} = 137{,}89 \text{ GPa}$

$E_{22} = E_{33} = 14{,}47 \text{ GPa}$

$G_{12} = G_{13} = G_{23} = 5{,}86 \text{ GPa}$

$\nu_{12} = \nu_{13} = \nu_{23} = 0{,}21$

**Fig. 1:** Schéma du quadricouche $(0°,90°)_S$ en traction de Wang et Crossman (1977)
*Fig. 1 : $(0°,90°)_s$ Wang and Crossman's laminate in tension (1977)*

Les solutions sont calculées analytiquement pour les modèles M4_5n et M4_(2n+1)M (Chabot, 1997). Pour cela, on utilise l'invariance par translation suivant x des champs de déformation. La résolution semi-analytique ainsi que le tracé des courbes sont effectués par le logiciel de calcul formel MATHEMATICA (version 3.0).

Présentons, par exemple, les résultats à l'interface 1, 2:

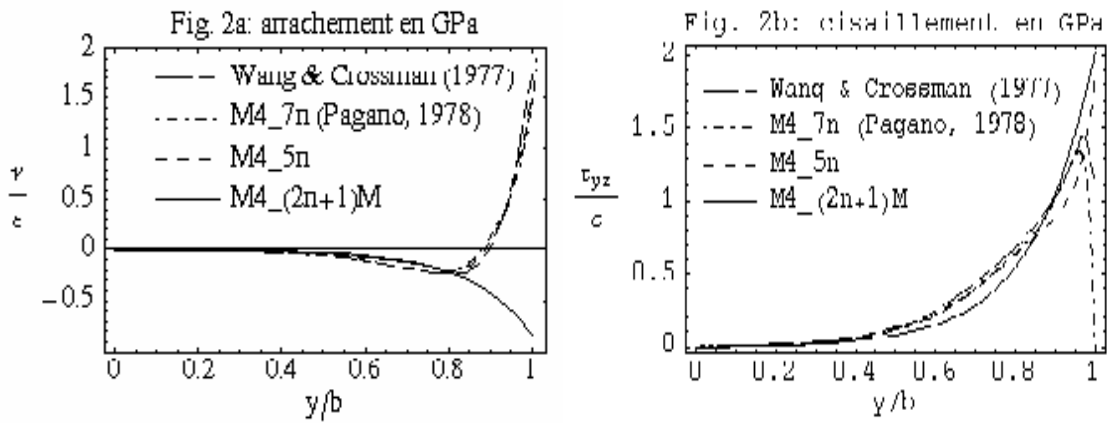

**Fig. 2:** Distribution des contraintes d'interface le long de l'interface 0°/90°
*Fig. 2 : 0°/90° interface stresses*

Si l'on compare nos résultats avec ceux du modèle local de Pagano (1978) ou du calcul par éléments finis 3D de Wang et Crossman (1977), on constate que la prévision de la distribution d'arrachement à l'interface 0°/90° (fig. 2a) est bonne pour le modèle M4_5n. Pour le modèle M4_(2n+1)M, l'arrachement à cette interface est obtenu par le calcul de la contrainte tridimensionnelle approchée $\sigma_{33}$ à la côte $z = h_1^+$. La résultante de cette contrainte devant être nulle, il faut compléter la distribution de $\sigma_{33}$ à l'interface par un effort d'arrachement concentré de type Dirac à l'extrémité $y = b$. Une preuve plus convaincante peut être trouvé dans Chabot (1997).

Ceci revient à considérer (fig. 2b) que la courbe de cisaillement du modèle M4_(2n+1)M admet en fait une discontinuité à l'extrémité avec $\tilde{\tau}.\underline{n} = 0$ au bord (sur cet interface l'effort d'arrachement est relié à la divergence de l'effort de cisaillement). Cet effort linéique concentré de type Dirac, dont l'intensité est relié au maximum des cisaillements au bord, pourrait être le support conceptuel naturel d'un critère de délaminage simple pour le dimensionnement de structure en matériau multicouche. Il peut être relié à la résultante des efforts d'arrachement à l'interface sur une distance « caractéristique » que l'on peut obtenir à l'aide de modèles plus fins d'où une similitude avec des critères de type contrainte moyenne.

## PLAQUES $(\theta_1, \theta_2)_s$ EN TRACTION

Nous. choisissons d'appliquer nos modèles multiparticulaires sur deux quadricouches en traction : un $(\pm \theta)_s$ et un $(\theta, \theta - 90°)_s$. Les systèmes d'équations sont résolus analytiquement par MATHEMATICA. Les données et le matériau sont les mêmes que précédemment. Respectivement pour les interfaces $\theta/-\theta$ et $\theta/\theta$-90° les plus sollicités, nous trouvons (Chabot, 1997) que si tôt que $\theta$ s'éloigne des angles à 0° et à 90° la distribution prépondérante d'efforts à l'interface est celle du cisaillement $\tau_{xz}$. Il est maximum au bord. Les deux courbes présentées ci-dessous donnent cette distribution de contrainte en fonction de l'angle $\theta$ pour le $(\pm \theta)_s$ (fig. 3a) et le $(\theta, \theta - 90°)_s$ (fig. 3b). Nous les avons normalisées par rapport à leur valeur maximum. La deuxième courbe prépare l'analyse des contraintes d'interface d'une plaque $(0°, 90°)_s$ trouée en traction, elle est comparée avec une analyse fine de type éléments finis menée par Raju et Crews (1981).

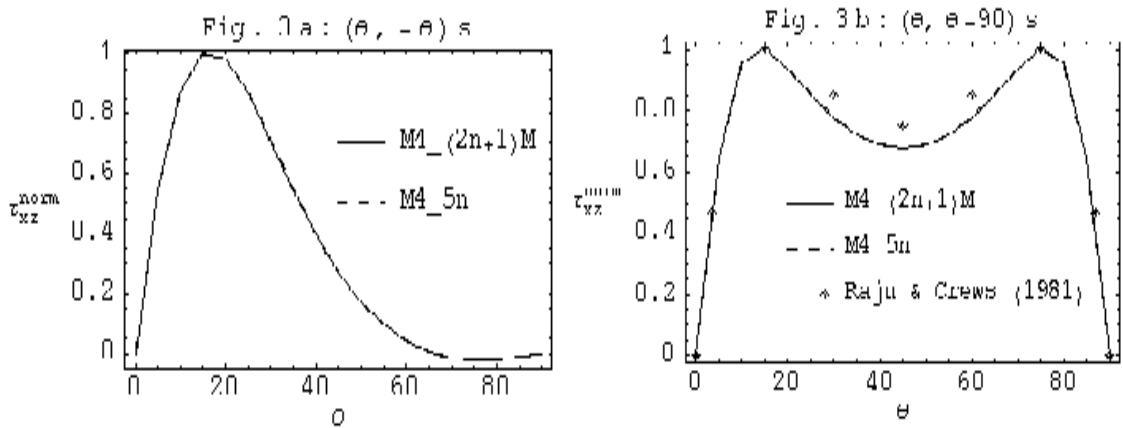

**Fig. 3:** Distributions des contraintes $\tau_{xz}$ maximum au bord
*Fig. 3 : Maximum edge $\tau_{xz}$ value*

Les angles donnant le maximum des cisaillements sont bien prédits par nos modèles.

### EXEMPLE DE LA PLAQUE $(0°, 90°)_s$ TROUEE EN TRACTION

Dans le cas où le rayon du trou est supérieur à trois fois l'épaisseur totale de la plaque, Raju et Crews (1982) ont montré par une analyse éléments finis très fine que l'on peut prévoir simplement les concentrations de contrainte au bord du trou. Pour ce faire, il suffit de savoir analyser les solutions de la plaque $(\theta, \theta - 90°)_s$ non trouée en traction que l'on combine à la déformation $\varepsilon_{\theta\theta}$ d'une plaque homogène équivalente calculée analytiquement par la méthode de Lekhnitskii (1968) au bord du trou (Cf. fig. 4a), p étant le chargement de cette plaque. Nos modèles multiparticulaires donnent facilement la distribution prépondérante des efforts de cisaillement $\tau_{\theta z}$ pour l'interface 0°/90° (Cf. fig. 4b où l'angle est compris entre l'axe perpendiculaire à la direction de traction et cette même direction).

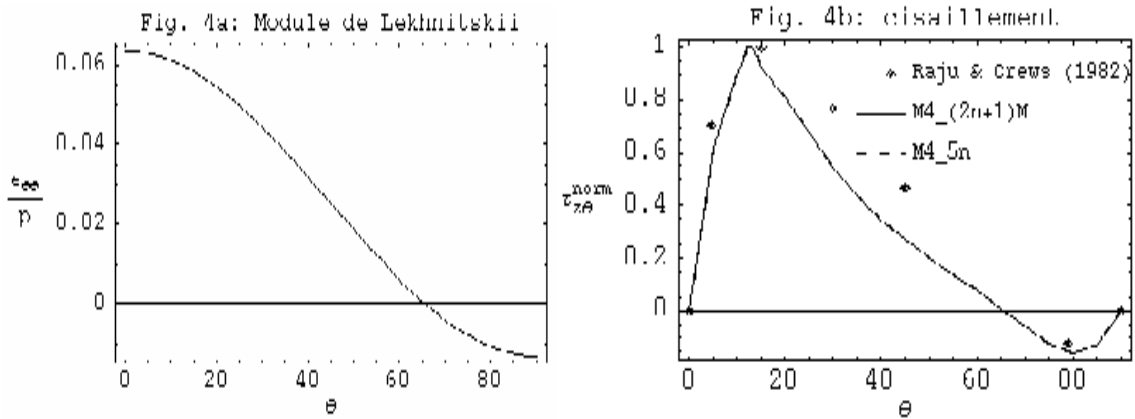

**Fig. 4:** Courbes de la plaque trouée $(0°, 90°)_s$ en traction

*Fig. 4 : curves of $(0°, 90°)_s$ stacking with a hole*

On observe ainsi une bonne concordance des résultats donnés par les modèles M4_5n, M4_(2n+1)M et l'analyse éléments finis de Raju et Crews (1982). Ils prévoient un pic de cisaillement aux alentours de 75° par rapport à la direction de traction. Ceci est confirmé par l'analyse éléments finis de Carreira (1998) où l'on montre que les solutions du modèle M4_5n sont très proches des solutions 3D. Le modèle M4_(2n+1)M, malgré ses équations sommaires est un indicateur pertinent, très simple d'utilisation, des phénomènes 3D.

## CONCLUSION

En reprenant l'approche développée par Pagano dans son modèle "local" (1978) et en y introduisant certaines simplifications, nous avons construit une famille de Modèles Multiparticulaires des Matériaux Multicouches. Dans ce papier, nous en présentons deux : le model M4_5n, qui s'apparente à un empilement de n plaques de Reissner-Mindlin et le model M4_(2n+1)M, qui est une généralisation du Shear Lag Analysis. Ces modèles se prêtent à de nombreuses solutions analytiques. Nous avons présenté la prévision des effets de bord par ces modèles dans la traction d'empilement $(0, 90°)_s$. $(\pm \theta)_s$ $(\theta, \theta - 90°)_s$ et d'une plaque $(0, 90°)_s$ trouée. Le modèle M4_5n donne sur ces problèmes une bonne prédiction des champs de contraintes 3D. Il peut être facilement implémenté dans un code de calcul classique disposant d'éléments de type Reissner-Mindlin. Le modèle M4_(2n+1)M ne permet pas une prédiction fidèle des champs de contraintes 3D mais il présente l'avantage de donner une information synthétique en faisant apparaître un effort concentré au bord. Cet effort concentré pourrait être le support conceptuel naturel d'un critère de délaminage simple pour le dimensionnement de structure en matériau multicouche similaire, dans l'esprit, aux critères en contrainte moyenne

## BIBLIOGRAPHIE